\documentclass[12pt]{iopart}

\usepackage{amssymb}
\usepackage{color}
\usepackage{graphicx}

\def\mylabel#1{\label{#1}}

\def\AA{$\mathring{\rm A}\,$}

\def\p{\mat p}
\def\s{\mat s}
\def\x{\mat x}

\def\balpha{{\bar{\alpha}}}

\def\<{\langle}
\def\>{\rangle}
\def\[{\left[}
\def\]{\right]}
\def\({\left(}
\def\){\right)}

\def\re{ {\rm Re} }
\def\im{ {\rm Im} }

\def\a0{~$a_0$}

\def\TWcm{~TW/cm$^{2}$}

\begin{document}
\title[]{
Coalescence of two branch points in complex time marks the end of
rapid adiabatic passage and the start of Rabi oscillations
}

\author{
Petra Ruth Kapr\'alov\'a-\v Z\v d\'ansk\'a$^{1}$,
Milan \v Sindelka$^{2}$,
Nimrod Moiseyev$^{3,4}$
}

\address{$^1$ Department of Radiation and Chemical Physics,
Institute of Physics, Academy of Sciences of the Czech Republic,
Na Slovance 2, 182 21 Prague 8, Czech Republic}
\ead{kapralova@fzu.cz}

\address{$^2$ Institute of Plasma Physics, Academy of Sciences of the Czech
Republic, Za Slovankou 1782/3, 182 00 Prague 8, Czech Republic}
\ead{sindelka@ipp.cas.cz}

\address{$^3$ Schulich Faculty of Chemistry, Technion -- Israel Institute of Technology, Haifa, 32000, Israel}
\address{$^4$ Faculty of Physics, Technion -- Israel Institute of Technology, Haifa, 32000, Israel}
\ead{nimrod@tx.technion.ac.il}

%\ams{}
\submitto{\jpa}

\begin{abstract}	
We study theoretically the population transfer in two-level atoms driven by chirped lasers.
It is known that in the Hermitian case,
the rapid adiabatic passage (RAP) is stable for an above-critical chirp
below which the final populations of states Rabi oscillate with varying laser power.
We show that if the excited state is represented by a resonance, the
separatrix marking this critical phenomenon in the space of the laser pulse parameters 
emanates from an exceptional point (EP) -- a non-Hermitian singularity formed in the atomic system by the fast laser field oscillations
and encircled due to slow variations of the laser pulse envelope and instantaneous frequency. 
This critical phenomenon is neatly understood via extending the ``slow'' time variable into the complex plane,
uncovering a set of branch points which encode non-adiabatic dynamics, where
the switch between RAP and Rabi oscillations is triggered by a coalescence of two such branch points.
We assert that the intriguing interrelation between the two different singularities -- the EP
and the branch point coalescence in complex time plane -- can motivate feasible experiments involving laser driven atoms.
\end{abstract}

\maketitle

\section{Introduction}

\noindent
Exceptional points (EPs) represent non-Hermitian degeneracies of a Hamiltonian
which form branch points in the plane of external Hamiltonian parameters~\cite{Kato,Berry:2004,Moiseyev}.
The non-Hermitian degeneracy is very different from the degeneracy in Hermitian Hamiltonians since 
the two corresponding eigenvectors coalesce and a self-orthogonal state is obtained. 
When an EP is stroboscopically encircled in the parameter plane, states are
exchanged as the complex energies form Riemann surfaces.
This of course sparkled interest in quantum dynamics of a system
which is forced to encircle EP by changing Hamiltonian parameters in time,
a process called {\it dynamical encircling} and primarily addressed in the present paper.

Exceptional points (EPs) have been initially used as a
mathematical concept in quantum theory~\cite{HEISS:1970,MOISEYEV:1980}.
Their physical significance was first recognized by Berry in 1994 based
on an early observation by Pancheratnam in 1955~\cite{BERRY:1994,Zienau:1976}.
A topological structure as a basic phenomenon related to EPs was first predicted in 1998~\cite{Heiss:1998}
and verified experimentally in 2001 and 2003~\cite{Dembowski:2001,Dembowski:2003}.
Since then EPs have been studied in many papers, see reviews~\cite{Rotter:2009,Rotter:2015,Heiss:2012,Miri:2019}.
A particular interest in the EP physics arose after 2008, when an occurence of EP
has been associated with an explanation for the
breakdown of the real spectrum of parity-time (PT) 
symmetric non-Hermitian Hamiltonian 
(see Refs.~\cite{Bender:1998,Bender:2007}),
which takes place in two coupled gain and loss waveguides~\cite{Klaiman:2008}.
Many experimental demonstrations of EPs in optical devices~\cite{Regensburger:2012,Feng:2013} and laser cavities~\cite{Liertzer:2012,Peng:2014,Feng:2014,Ozdemir:2019} followed.  
Though less often, EP phenomena in atomic and molecular physics have also been measured or proposed theoretically, 
notably for PT-symmetric systems designed using atomic vapor cells and optical traps~\cite{Peng:2016,Li:2019},
and for isolated atoms and molecules perturbed by fields~\cite{Cederbaum:2011,Atabek:2011,Leclerc:2017,Benda:2018,Pick:2019}.

The present study follows up on previous theoretical studies of non-adiabatic dynamics which
occurs when {\it dynamically encircling the EP} by a
contour in the plane of the relevant Hamiltonian
parameters. Current literature on this problem has
been mainly focused on the state switching phenomenon, where the non-Hermitian components
of the energies, preventing the usual application
of the adiabatic theorem, induce a chiral behavior~\cite{Uzdin:2011,Gilary:2013,Graefe:2013,Kapralova-Zdanska:2014}. Recently, experimental realizations of the
time-asymmetric switch phenomena were reported
in waveguides~\cite{Doppler:2016,Xu:2016}.
As a matter of fact, it
would be difficult to tailor a similar experiment in
the scope of atomic and molecular physics. 

Here
we present a different phenomenon
represented by a behavior switch between Rabi oscillations and rapid
adiabatic passage, which
is attributed to be inherently associated with the EP
encircling, see Section~2. This phenomenon is even particularly
suitable for an experimental realization in laser
driven atoms which is the objective of Sections~3 and 4.
In Section~3 we present a unique
experimental strategy to determine the behavior switch, which as we will
see is
not a straightforward thing due to small variations of the measured
quantity.
In Section~4 we enlist and discuss important requirements on
suitable laser--atom systems, demonstrating the behavior switch in
a full-dimensional simulation 
of the driven helium atom.
After presenting Conclusions we append a theoretical derivation
of transformation between frequency and time domains for chirped
laser pulses which based on Wigner phase space theory and used
within this work.

\section{Behavior switch between Rabi oscillations and rapid adiabatic passage}

\subsection{Dynamical EP encircling resulting in the behavior switch}

\noindent
Let us motivate our considerations by recalling
the well known problem, where two bound states
of an atom are coupled by a continuous wave laser.
Such atoms exhibit {\it Rabi oscillations}, which appear also in the more general case of a finite pulse, where the populations of the states oscillate with
integer multiples of $\pi$ of the pulse area~\cite{Vitanov:2001}. On
the other hand, a completely different kind of dynamics takes place whenever a (large enough) frequency chirp is added, which leads to the rapid adiabatic passage (RAP)~\cite{Vitanov:2001,Tannor}. These facts indicate
that some kind of a switch between the monotonic
(RAP) and oscillatory (Rabi) behavior must occur
at some critical chirp. This assumption is strongly
supported by the known instability of RAP for insufficient chirping~\cite{Malinovsky:2001}.
The aim of our present
work is to explore the switch between the regimes
of Rabi oscillations and RAP in the more general
case, when the upper atomic level corresponds to
a metastable resonance state. A clear cut (yet
so far unreported) link between the Rabi-to-RAP
switch and the EP-encircling dynamics will be subsequently highlighted.

Our Gaussian chirped pulse is defined in the time-domain by
the laser strength $\varepsilon_0(t)$ and instantaneous frequency $\omega(t)$,
such that
\begin{eqnarray}
	& \varepsilon_0(t) = \varepsilon_{0}^{max} e^{-(t/\tau)^2/2}, 
\quad \omega(t) = \omega_r + \alpha t .
	\mylabel{EQ2}
\end{eqnarray}
The two-level atom consists of one bound and one metastable state, where
the transition frequency is $\omega_r$, 
ionization width of the metastable state is $\Gamma$, and
the transition dipole moment is $\mu$.
We restrict our discussion to the systems where $\mu$ is real valued.
The driven atomic system is described in the rotating wave approximation (RWA)~\cite{AllenEberly}
by the interaction Hamiltonian 
\begin{eqnarray}
H(t) = \hbar \[
\matrix{
0 & \frac{1}{2}\Omega(t) \cr
 \frac{1}{2}\Omega(t) & \Delta(t)
}
\],
\mylabel{EQ1}
\end{eqnarray}
where the time-dependence is related to the slow variation
of the field frequency and amplitude within the laser pulse.
$\Delta(t)=\omega(t) - \omega_r +i\Gamma/2\hbar$ stands for the dynamical frequency
detuning;
and $\Omega(t)=\mu\varepsilon_0(t)/\hbar$ is the time-dependent Rabi frequency.

Dynamical quantum state of our laser driven atom can be then described by the formula
\begin{eqnarray}
	&\psi(t) = e^{-\frac{i}{\hbar}\int\limits_{0}^t d t' \epsilon_-(t')} a_-(t) \Phi_-(t) 
	\cr&\quad + e^{-\frac{i}{\hbar}\int\limits_{0}^t d t' \epsilon_+(t')} a_+(t) \Phi_+(t) ,
\mylabel{EQ3a}
\end{eqnarray}
where $\Phi_\pm(t)$ are the two parametrically $t$-dependent eigenvectors of the Hamiltonian (\ref{EQ1}), 
associated with complex eigenvalues $\epsilon_\pm(t)$,
\begin{eqnarray}
\epsilon_\pm(t) = \frac{\hbar}{2}\[\Delta(t)\pm\sqrt{\Delta^2(t)+\Omega^2(t)}\].
\mylabel{EQ3}
\end{eqnarray}
If the time-dependence is omitted in Eq.~\ref{EQ1}, then such a Hamiltonian would
describe an atom driven by a continuous wave laser, with parameters $[\omega,\varepsilon_0]$.
This Hamiltonian has an EP where $\epsilon_+=\epsilon_-$ for
$\omega^{EP}=\omega_r$ and $\varepsilon_{0}^{EP}=\Gamma/2\mu$.  
If $t$ is taken as an adiabatic parameter in Eq.~\ref{EQ1}, the EP 
is encircled by trajectories $[\omega(t), \varepsilon_0(t)]$ defined by the pulse of Eq.~\ref{EQ2}
for $\varepsilon_0^{max}>\varepsilon_0^{EP}$, see Fig~1.
We assign $\Phi_-(t\to-\infty)$ to represent the field free bound state, which is
initially populated, $a_-(t\to -\infty)=1$ and $a_+(t\to -\infty)=0$.
Counterintuitively,
the bound state at the end of the pulse is represented by
$\Phi_+(t\to\infty)$ due to the adiabatic flip~\cite{Lefebvre:2009}. 
Accordingly,
the final population (survival probability)
of the bound state $p_b$ is given by the ratio of the coefficients in the front of $\Phi_+(t\to\infty)$ 
and $\Phi_-(t\to-\infty)$ in Eq.~\ref{EQ3a}, such that
\begin{eqnarray}
	p_b = \left |
	a_+(t\to\infty) \cdot e^{-\frac{i}{\hbar} \int\limits_0^\infty dt'\,\epsilon_+(t')}
	e^{\frac{i}{\hbar} \int\limits_0^{-\infty} dt'\,\epsilon_-(t')}
	\right |^2 .
	\mylabel{EQPT}
\end{eqnarray}
\begin{figure}[h!]
	\begin{center}
		\includegraphics[width=3.5 in]{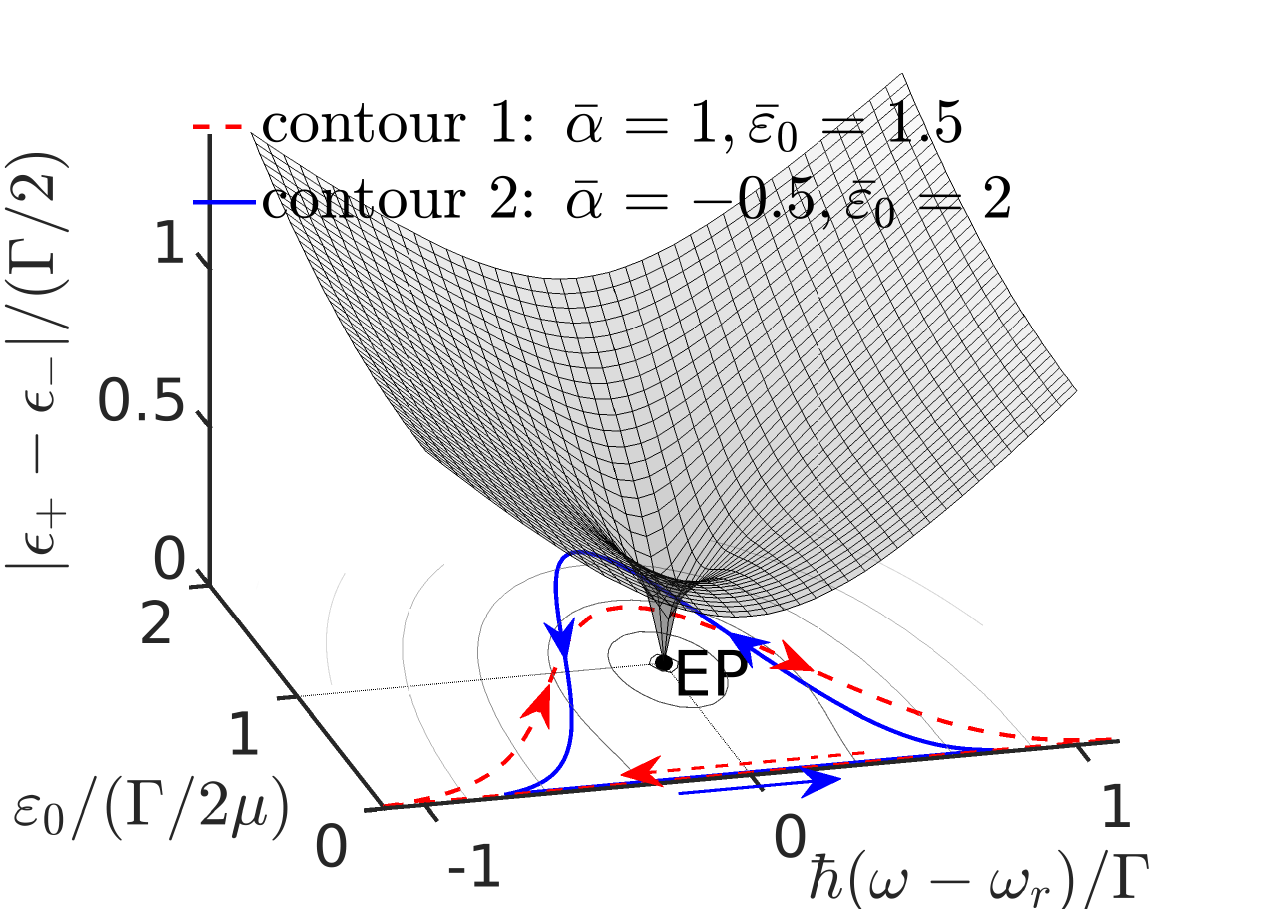}
	\end{center}
\caption{
	An exceptional point (EP) is formed in a two-level 
	atom including bound and resonance states (width $\Gamma$)
	driven by a continuous wave laser
	for the well defined laser parameters $[\varepsilon_0=2\mu/\Gamma,\omega=\omega_r]$
	assuming $\mu\in\Re$.
	Gaussian chirped laser pulses define adiabatic encircling contours, which vary for
	different values of the chirp $\alpha$ and peak strength $\varepsilon_0^{max}$. 
	The plot is based on system independent parameters, $\bar \alpha$ and
	$\bar\varepsilon_0$ are defined below in Eq.~\ref{barpar}.
}
\end{figure}

The non-adiabatic amplitudes $a_\pm(t)$ of Eq.~\ref{EQ3a} satisfy the close-coupled equations
\def\DE#1{[\epsilon_-(#1) - \epsilon_+(#1)]}
\begin{eqnarray}
&\dot a_+(t) = -a_-(t) N(t) e^{-\frac{i}{\hbar}\int\limits_{0}^{t} dt' \DE{t'}} \cr
	&\dot a_-(t) = a_+(t) N(t) e^{\frac{i}{\hbar}\int\limits_{0}^{t} dt' \DE{t'}}  ,
\mylabel{EQN4}
\end{eqnarray}
where $N(t)$ is the non-adiabatic coupling defined as $(\Phi_+|\partial_t \Phi_-)$, 
where the non-Hermitian $c$-product is used~\cite{Moiseyev}.

The adiabatic energies $\epsilon_\pm$ in Eq.~\ref{EQN4} are complex-valued, which 
causes a breakdown of the adiabatic theorem~\cite{Uzdin:2011,Gilary:2013,Graefe:2013,Kapralova-Zdanska:2014}.
Yet, if the initial state is the less dissipative one and the
excited state is the more dissipative one, 
an adiabatic perturbation theory is 
justified to solve Eq.~\ref{EQN4} and express the 
non-adiabatic amplitudes for the encircling of EP~\cite{Dridi:2010}.
Thus the underlying assumption for our present discussion
is given by,
\begin{equation}
	a_+(t\to\infty) = - \int\limits_{-\infty}^\infty dt\, N(t) e^{\frac{i}{\hbar}\int\limits_{-\infty}^{t} dt' \DE{t'} } .
\mylabel{EQ6}
\end{equation}
See our paper~\cite{PAPER} for a detailed derivation of
Eq.~\ref{EQ6} and a thorough numerical verification of the convergence of
the perturbation series for the system under the study here.
The survival probability $p_b$ (Eq.~\ref{EQPT})
is calculated by numerically evaluating the integral Eq.~\ref{EQ6}
for different values of the pulse parameters. 
We conveniently use effective laser parameters 
proposed in Ref.~\cite{PAPER}, which allow to describe laser induced dynamics for different atomic transitions
as a single atom-independent problem. Namely, we set
\begin{eqnarray}
	& \theta = \tau\varepsilon_0^{max} \mu \sqrt{2\pi},
	\quad
	\bar \varepsilon_0 = \frac{\varepsilon_0^{max}}{\varepsilon_0^{EP}} = \frac{2\mu\varepsilon_0^{max}}{\Gamma}, \cr
	&
	\bar \alpha = \frac{\alpha\tau}{\varepsilon_0^{max}} \frac{2\hbar}{\mu},
	\mylabel{barpar}
\end{eqnarray}
where
$\theta$ defines the temporal pulse area~\cite{Vitanov:2001},
$\bar\varepsilon_0$ is the peak strength relative to the position of EP,
$\varepsilon_0^{EP} = \Gamma/2\mu$, 
and $\bar\alpha$ represents an effective chirp.

Importantly, Fig.~\ref{F1} shows that
$p_b$ creates a unique pattern in the $[\bar\varepsilon_0,\bar\alpha]$ plane,
where separate areas of oscillatory and monotonous behavior appear.
The separatrix $s$ demarcates between the parametric regions
associated with the dynamical regimes of Rabi oscillations and RAP. 
This observation highlights an emergence of the critical phenomenon
associated with dynamical encircling of the EP, and the main topic of this paper.
At large values of $\bar{\varepsilon}_0$, the separatrix is nearly parallel to the $\bar{\varepsilon}_0$-axis.
However, as $\bar{\varepsilon}_0$ decreases, the separatrix becomes sharply bent
towards the ${\bar{\varepsilon}_0}$-axis, and intersects this axis at
 the peak strength 
$\bar\varepsilon_0=1$, which represents the position of the encircled EP,
$\varepsilon_0^{max}=\Gamma/2\mu$.
Note that the position of the separatrix is not changed for different values of the pulse area $\theta$,
but with larger $\theta$, the density of oscillations is increased whereas
the average value of $p_b$ is rapidly decreased.

\begin{figure}[h!]
	\begin{center}
	\includegraphics[width=4 in]{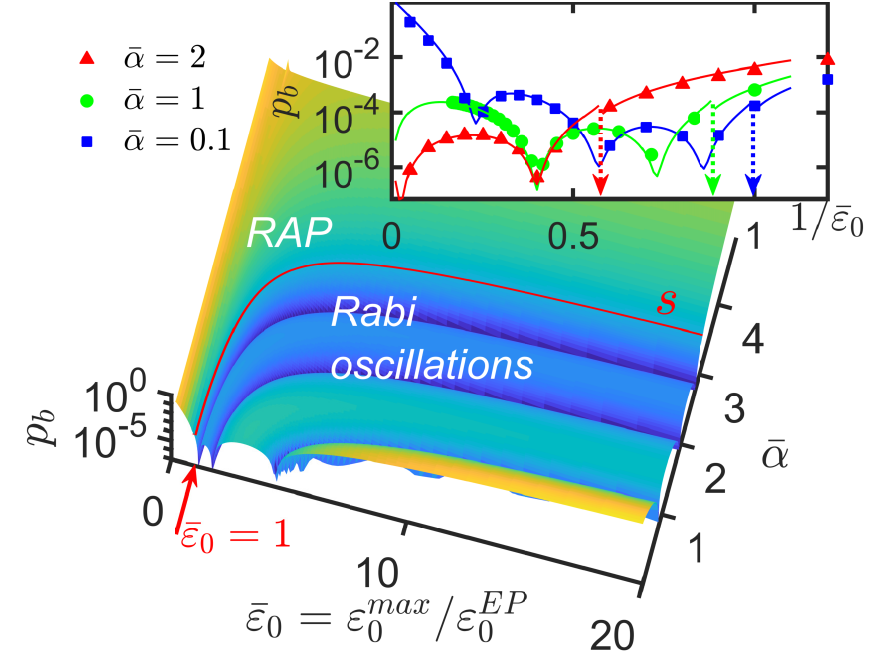}
\end{center}
\caption{
Survival probability $p_b$ of an initial bound state coupled to an unstable resonance by a chirped laser. 
The laser pulse is defined by effective parameters $\bar\varepsilon_0$, $\bar\alpha$, $\theta$ of Eq.~\ref{barpar}, which allow us to draw a
single plot for different atomic Hamiltonians.
(The pulse area $\theta$ is held fixed, $\theta=6\pi$).
The main plot, based on numerical results of the first-order perturbation approximation, Eq.~\ref{EQPT},
demonstrates an abrupt switch between the regimes of Rabi oscillations
and rapid adiabatic passage (RAP). The separatrix $s$ emanates from the laser amplitude
of the encircled EP ($\varepsilon_0^{max}=\varepsilon_0^{EP}$)
and for asymptotically large pulse amplitudes ($\varepsilon_0^{max}\gg\varepsilon_0^{EP}$)
defines a critical chirp for the Hermitian RAP (defined by $\Gamma,\varepsilon_0^{EP}\to0$).
The inset additionally shows the obtained analytical solution (lines, Eq.~\ref{EQ9}), 
compared with realistic full-dimensional simulations 
for the transition $^1P^o (1s2p)$ $\to$ $^1S^e (2p^2)$ in the helium atom (markers),
which are discussed in the end of the paper.
The position of the separatrix $s$ is denoted here by the dotted arrows.
}
\mylabel{F1}
\end{figure}

\subsection{Coalescence of branch points in complex time}

\noindent
Let us move on to a theoretical explanation 
of the above described critical phenomenon.
This will require 
moving to a complex plane of adiabatic time,
where the change of behavior of $p_b$ is related to a particular 
non-Hermitian singularity which is formed.
Let us start with a definition of
 the non-adiabatic coupling $N(t)$~\cite{Shore:2008}
\begin{equation}
N(t) =
\frac{1}{4i} \frac{d (\Omega/\Delta)}{dt}
\(
\frac{1}{i + \Omega/\Delta} +
\frac{1}{i - \Omega/\Delta}
\)
\mylabel{EQ7}
\end{equation}
showing that $N(t)$
includes poles $t_k$ in the complex plane of time $t$,
where $\Omega(t_k)=\pm i\Delta(t_k)$, $t_k\in C$.
The poles $t_k$ also represent complex degeneracies of 
the analytically continued Hamiltonian $H(t_k)$; 
this becomes clear by putting the relation
$\Omega(t_k)=\pm i\Delta(t_k)$ to Eq.~\ref{EQ3}.
Honoring the nomenclature established
by Dykhne, Davis and Pechukas, Child~\cite{Dykhne:1962, Davis:1975, Child:1978},
and others we shall refer to these points
as {\it transition points (TPs)}, though
the nature of these points is exactly same as of the EPs. We propose to reserve the latter term to the branch points if the 
Hamiltonian parameters are other than time.

Let us explore now in more detail the properties of the TPs, their intimate association with
the critical phenomenon shown in Fig.~2, and their relation to the EP. Fig.~3 shows the distribution of 
TPs in the complex time plane for the studied two-level system driven by the Gaussian linearly
chirped pulse of Eq.~\ref{EQ2}.
The TPs in the upper half of the complex time plane
are typically represented by a pair of points near the real axis and two 
asymptotic sequences of
other TPs which unfold within the complex plane at the angles of $\pm\pi/4$
 as a gradually condensing series of points, see Fig.~3.
We shall restrain here to a shorthand argument to explain such a distribution of the TPs.
The TPs represent zeros of the quasienergy split, namely the sum of 
$[\Omega^2(t)+\Delta^2(t)]$, see Eq.~\ref{EQ3}.
We seek points $t_k$ at which contributions of these two
terms cancel out. Based on this requirement
$\Omega^2(t_k)$ and $\Delta^2(t_k)$ must have an opposite complex phase.
Eqs.~\ref{EQ2} and \ref{EQ1} imply
\begin{eqnarray}
&\Omega^2(t_k) \propto \exp(-t_k^2/\tau^2), \cr
&\Delta^2(t_k) \propto (\alpha t_k + i\Gamma/2\hbar)^2.
\mylabel{EQTP0}
\end{eqnarray}
Hence, for large $|t_k|$, the term $[\Omega^2(t)+\Delta^2(t)]$ vanishes iff
\begin{eqnarray}
(t_k/\tau)^2 = k\cdot2i\pi \pm i\pi/2 + \beta,\quad \beta\in\Re ,\quad k\in Z .
\mylabel{EQTP1}
\end{eqnarray}
Clearly, both $\Omega^2(t_k)$ and $\Delta^2(t_k)$ possess purely imaginary values along the progression of the $t_k$'s.

Using the same argument we will show how
the central pair of TPs produces a specific singularity triggering
the critical phenomenon of Fig.~2.
Namely, let us assume small values of $|t_k|$, and set for convenience $z_k=-i t_k$.
Eqs.~\ref{EQTP0}   imply then
\begin{eqnarray}
&\Omega^2(i z_k) \propto 1+z_k^2/\tau^2+\cdots, \cr
& \Delta^2(i z_k) \propto -(\alpha z_k + \Gamma/2\hbar)^2 ,
\mylabel{EQTP00}
\end{eqnarray}
showing that $[\Omega^2(i z_k)+\Delta^2(i z_k)]$ represents approximately a quadratic polynomial.
The sought central pair of TPs  is determined by roots $z_{1,2}$ of this quadratic polynomial.
Depending upon the laser pulse parameters, these roots $z_{1,2}$ are located either at the real axis (blue triangles of Fig.~2),
or they penetrate into the complex $z$-plane (red bullets of Fig.~2). 
Critical transition from the real $z$-axis into the complex plane 
occurs then at those specific laser parameters for which $z_1=z_2=\tilde{z}$.
The degenerate case of $z_1=z_2=\tilde{z}$ corresponds actually to {\it coalescence} of two TPs, i.e., to merging of
$z_1$ and $z_2$ into a single TP $\tilde{z}$ of a higher Puiseux order, such that
$\epsilon_+(iz) - \epsilon_-(iz) = {\cal O}(z-\tilde{z})$,
where the leading square root term in the 
Puiseux series vanishes. 
(Note that this is still a second-order non-Hermitian degeneracy, as opposed to other 
situations where
a higher Riemann-order EP is formed~\cite{Am-Shallem:2015}).

\begin{figure}[h!]
	\begin{center}
	\includegraphics[width=4 in]{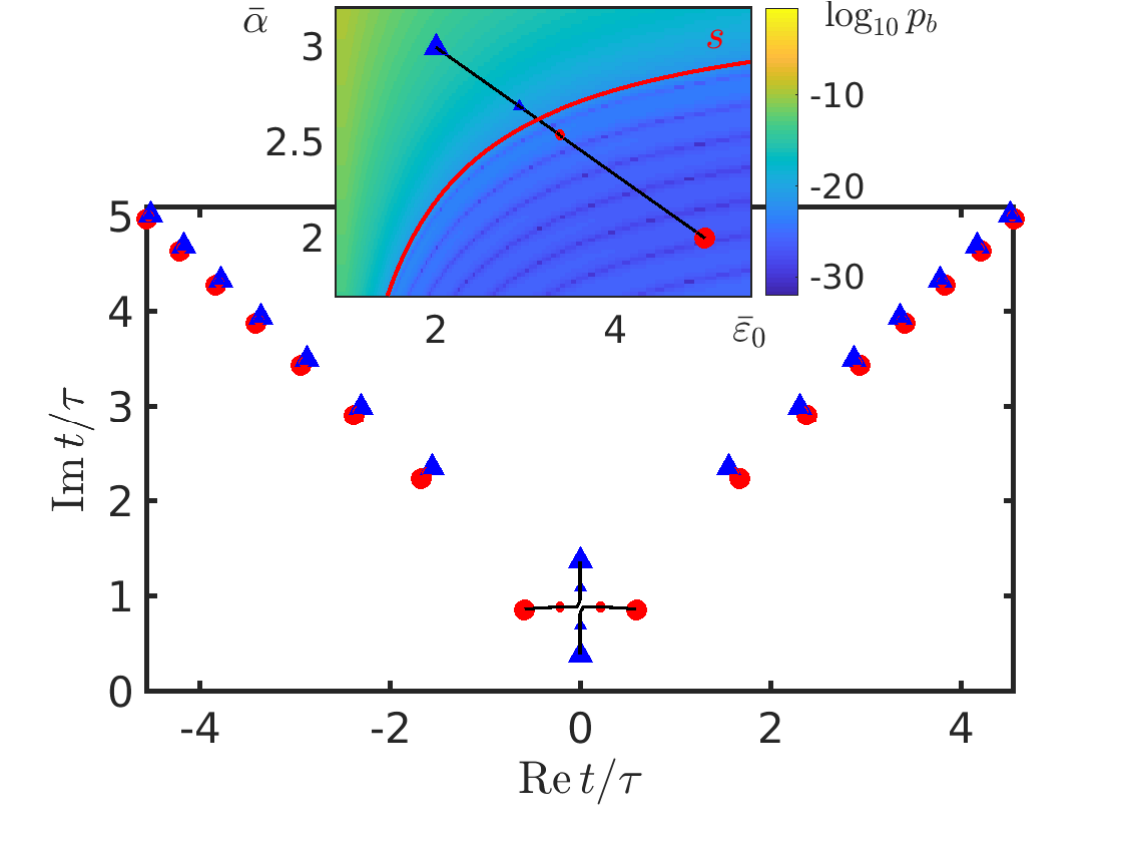}
	\end{center}
\caption{
TPs ($\blacktriangle$, $\bullet$)
arising in the adiabatic Hamiltonian (Eq.~\ref{EQ1}) analytically continued
to the complex time plane. The
positions of the TPs vary with the pulse parameters $\bar\alpha,\bar\varepsilon_0$ 
shown by corresponding markers in the inset.
Critical change of behavior of $p_b$ displayed in the inset for $\theta=30\pi$
is associated with the coalescence of the central TPs.
}
\end{figure}

Importantly, the just discussed TPs enable us to neatly explain
the behavior of the time integral of Eq.~\ref{EQ6} and its changes
with varying laser parameters. The corresponding argument is based upon contour integration in the plane of complex time~\cite{PAPER}.
Mathematical elaborations of Ref.~\cite{PAPER} provide ultimately
a simple outcome.  Namely, for a sufficiently large pulse area $\theta$,
the pair of TPs near the real axis outweighs residual contributions
of the other TPs and
actually controls the value of the integral of Eq.~\ref{EQ6}.
The fact that the
 central pair of TPs may acquire two different configurations in the complex time plane
  is of a fundamental
importance for an explanation of the studied critical phenomenon reported in Fig.~1. 
If the two TPs possess nonzero real parts (bullets in Fig.~3),
their residual contributions (given by complex exponentials)
sum up into a cosine term, while if they lay on the imaginary axis (triangles in Fig.~3)
they sum up to a single exponential term.
The survival probability of Eqs.~\ref{EQPT} and \ref{EQ6}
takes then the form \cite{PAPER}
\begin{eqnarray}
	&	p_b = \(\frac{\pi}{3}\)^2 \,\exp\[-\frac{\theta}{\hbar\sqrt{2\pi}}\,\frac{\bar \gamma}{\bar\varepsilon_0} \]
	\times Z(\theta,\bar\varepsilon_0,\bar\alpha), 
	\mylabel{EQ9}
\end{eqnarray}
where the term $Z$ acquires two distinct values for the two just mentioned configurations of the two central TPs.
Namely, if the TPs possess nonzero real parts, then
\def\x{\bar\varepsilon_0}
\def\aux{\frac{\theta}{\hbar\sqrt{2\pi}}\frac{\phi}{\x}}
\begin{eqnarray}
	&	Z=4\, \cos^2\[\aux\] , 
	\mylabel{EQ10}
\end{eqnarray}
but if they are located on the imaginary axis then 
$Z=1$.
$\phi$ and $\bar\gamma$ are associated with the residua at the central TPs and
must be obtained numerically; note that the residua and thus
also $\phi$ and $\bar\gamma$ do not depend on the prefactor $\theta$.
 The regularity of the oscillations displayed in Fig.~2 indicate that
$\phi/\bar\varepsilon_0$ and $\bar\gamma/\bar\varepsilon_0$ have certain important characteristics further studied in Section~\ref{S3}.
%Namely, if we define the plane of the laser parameters by $\bar\alpha$ and
%$1/\bar\varepsilon_0$ axes, and define a generalized radius on this plane as a measure of
%the distance from the coordinate center to the separator, then
%$\phi/\bar\varepsilon_0$ and $\bar\gamma/\bar\varepsilon_0$ are almost solely dependent on the radius.
%Within the oscillatory region,
%$\phi$ is approximated by a linearly increasing function
%with the radius, whereas $\bar \gamma/\bar\varepsilon_0$ is roughly parabolic with the maximum near the separator $s$.
%In the monotonic region, $\bar\gamma/\bar\varepsilon_0$
%approaches zero with one over the radius.

Eqs.~\ref{EQ9} and \ref{EQ10} represent the final output of our theoretical analysis of $p_b$, and enable us to rationalize the corresponding abrupt change
of the behavior of $p_b$ from the regime of Rabi oscillations to RAP, as plotted in Fig.~2.
This Rabi-to-RAP transition occurs when crossing the separatrix $s$,
which is defined by the {\it coalescence} of the two
central TPs, Fig.~3.

\section{Proposal of experimental realization of Rabi-to-RAP switch}

\subsection{Analytical fit for the oscillatory term in the survival probability}

\begin{figure}	
	\begin{center}
		\includegraphics[width=4 in]{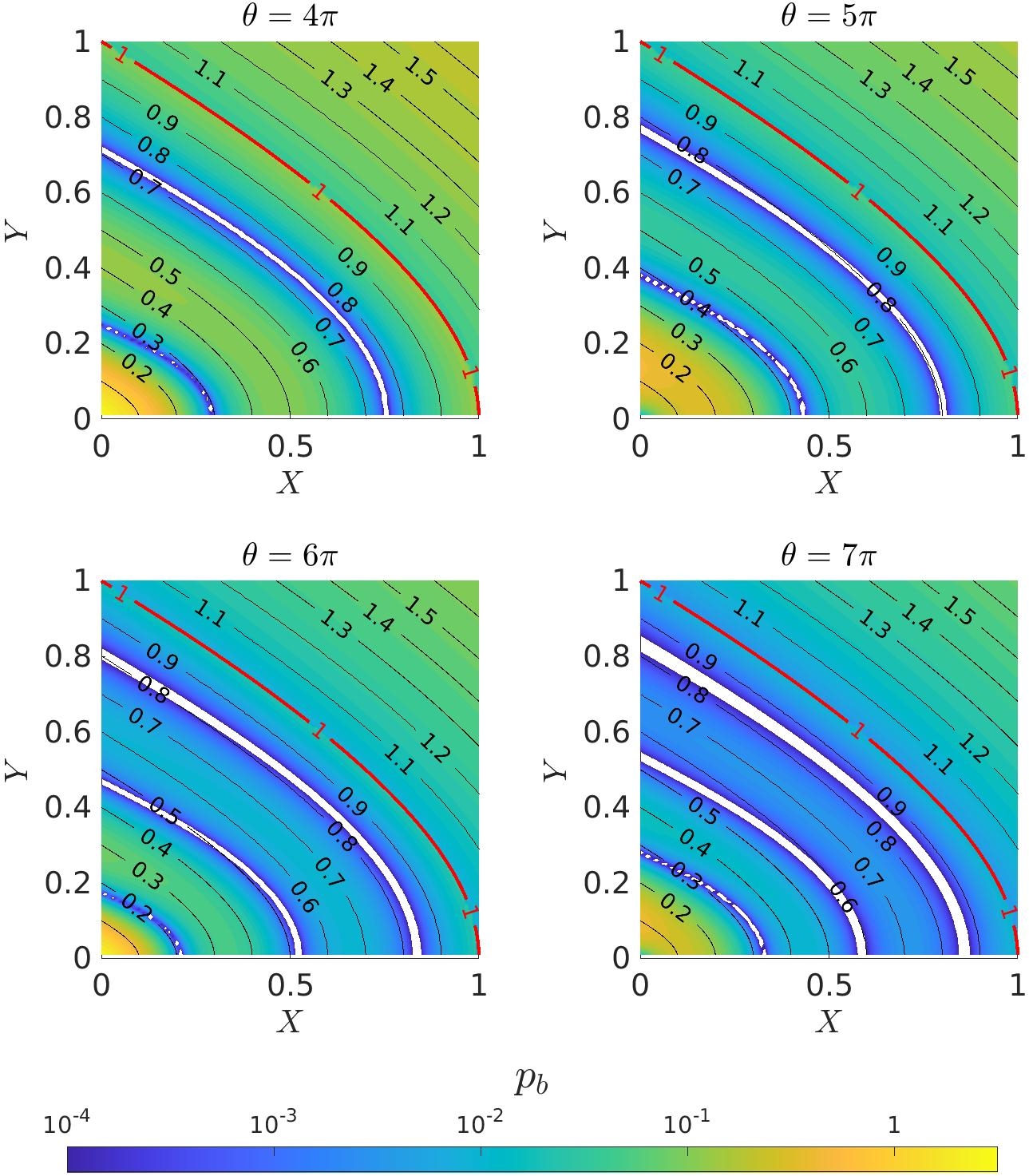}
	\end{center}
	\caption{
		The survival probability $p_b$ is displayed as a function
		of the reciprocal relative intensity ($X$-axis, $X=1/\bar\varepsilon_0$) 
		and effective chirp ($Y$-axis, $Y=\bar\alpha/2\sqrt{e}$), while the
		temporal pulse area $\theta$ is fixed on
		a given value (see above the frames).
		The values of the survival probability $p_b$
		are indicated by the
		shadow-scale (color-scale online) displayed in the 
		bottom.
		We define {\it generalized radius} $R$ as a relative distance
		from the coordinate center to the {\it separator}
		which occurs for $R=1$, see the labeled
		equivalue lines.
		This figure represents a numerical demonstration
		of the fact that the wavefronts of Rabi oscillations
		are approximately given by the equivalue lines of $R$.
	} 
	\mylabel{FigFrame1}
\end{figure}

\noindent
Let us provide a more detailed discussion of Rabi oscillations, for which the key is
 represented by the function $\phi/\bar\varepsilon_0$, see Eq.~\ref{EQ10}, where
$\phi$ is obtained via numerical integration, 
\begin{eqnarray}
&\phi = \frac{2\hbar}{\Gamma}\ \re \,
\int\limits_0^{t_0/\tau} ds\ \[\epsilon_-(s\tau) - \epsilon_+(s\tau) \]\, ,
\end{eqnarray}
where $t_0$ represents one of the central TPs, $t_0\in \cal C$,
and is found also numerically. Details of such numerical calculations 
are beyond the scope of the present report but they can be
found in Ref.~\cite{PAPER}, where it is also shown that they
yield the following approximate formula 
\begin{eqnarray}
	\frac{1}{\sqrt{2\pi}} \, \frac{\phi}{\bar\varepsilon_0}
	\approx
	-\frac{1}{2}\, (1-R) \, ,
	\mylabel{phi1}
\end{eqnarray}
where $R$ is defined as
\begin{equation}
	R = e^{-A^2/2} \cdot \( Y \sqrt{e} \, A + X \) ,
	\mylabel{EQ20a} 
\end{equation}
where
\begin{eqnarray}
	& X=\frac{1}{\bar{\varepsilon}_0},\quad
	& Y=\frac{\bar{\alpha}}{2\sqrt{e}} ,
	\mylabel{EQ20b}
\end{eqnarray}
and
\begin{eqnarray}
	& A = \frac{1}{2\sqrt{e}} \ \( \sqrt{\(\tan\Phi\)^2 + 4e} - \tan\Phi \) ,\quad
	& \Phi=\arctan \frac{X}{Y}\ .
	\mylabel{EQ20}
\end{eqnarray}

$R$ represents an effective distance,  a generalized radius,
on the $[X,Y]$-plane from the origin $[X=0,Y=0]$, where
the condition
\begin{eqnarray}
	R = 1 ,
	\mylabel{EQC}
\end{eqnarray}
defines the separator $s$, wherefore 
analytical expression for $R$ in Eq.~\ref{EQ20a}-\ref{EQ20}
is found as coalescence of zeros (TPs)
of the quasi energy split for Gaussian linear chirp.

We provide here a numerical evidence 
of the applicability of Eq.~\ref{phi1} in
Fig.~\ref{FigFrame1} which displays
survival probability $p_b$ as function of $X$ and $Y$.
Namely it is shown that wavefronts of Rabi oscillations 
approximately coincide with equivalue lines of $R$, while the period
of oscillations is approximately constant on the axis of $R$.

Eq.~\ref{phi1} allows us to obtain an approximate
analytical expression for the term $Z$ defining
Rabi oscillations in Eqs.~\ref{EQ9}-\ref{EQ10} such that 
\begin{eqnarray}
	&	Z\propto \cos^2\(\frac{\theta}{\hbar}\cdot \frac{R-1}{2}\) .
	\mylabel{EQZ}
\end{eqnarray}
This equation allows us to estimate
behavior of oscillations when different pulse parameters
are varied as we do in our considerations below.

\subsection{Using infinite pulse area limit to find the separator\label{S3} }

\noindent
In this Section we study behavior of Rabi oscillations
as function of temporal pulse area~$\theta$.
First we compare different panels of Fig.~\ref{FigFrame1}
which illustrate the effect of a gradual increase of the temporal pulse area.
The nodes (white spaces in the graphs) get more condensed with $\theta$ being increased. The node closest to the separator $s$ seems to approach 
and merge with the separator line the limit $\theta\to\infty$.
 Below we give a prove that this is indeed the case, and even that the
 same applies to the other subsequent nodes, which all drift
 towards the separator $s$ along with increasing $\theta$.

In order to study oscillations of survival probability $p_b$ as 
function of the temporal pulse area $\theta$ we choose a 
two-dimensional frame of laser pulse parameters defined by $[x,y]$ such that 
\begin{eqnarray}
	&x= \frac{\sqrt{2\pi} \hbar }{\theta}\ ,\quad
	&y= \frac{\balpha}{2}, \quad \delta_\omega = const.
	\mylabel{EQxy}
\end{eqnarray}
where $\delta_\omega$  represents a {\it frequency width of the laser pulse}. 
$\delta_\omega$ is used here to fully define 
the chirped laser pulse instead of the laser strength $\bar\varepsilon_0$,
see Eq.~\ref{EQ2}.
$\delta_\omega$ has the meaning of frequency
interval obtained when the chirped laser pulse defined
in Eq.~\ref{EQ2} is Fourier
transformed from the time to energy domain, see Appendix.
$\delta_\omega$ is related to $\bar\varepsilon_0$ through the
equation
\begin{eqnarray}
	\(
	\frac{\delta_\omega}{\Gamma} \frac{2}{\bar \varepsilon_0}
	\)^2
	=  
	2\pi
	\(\frac{\hbar }{\theta}
	\)^{2}+ \(\frac{\balpha}{2}
	\)^{2} \,  ,
\end{eqnarray}
which shows that the laser pulse intensity $\bar\varepsilon_0$
is represented by the reciprocal radius $1/r$ in the coordinate system $[x,y]$
such that
\begin{eqnarray}
\bar \varepsilon_0 =  \frac{2\delta_\omega}{\Gamma} \cdot \frac{1}{r}\, ,
\quad
r=\sqrt{x^2 + y^2}\ .
\mylabel{EQr}
\end{eqnarray}

Survival probability within coordinates $[x,y]$ is displayed in 
Fig.~\ref{FigFrame2}, where different panels correspond
to different frequency widths $\delta_\omega$.
The figures show  
that nodes of Rabi oscillations converge into a single point $[x=0,y=y_0]$,
where this phenomenon takes place for every possible value of $\delta_\omega$.
The crossing of nodal lines occurs for $x=0$ which represents the 
asymptotic limit of
the temporal pulse area $\theta$, see Eq.~\ref{EQxy}.
This fact partially supports our previous considerations
based on Figs.~\ref{FigFrame1} where we assumed that
the lines of nodes {\it merge with the separator}  
at the limit $\theta\to \infty$.
Let us prove that the crossing point $[0,y_0]$ 
is indeed found on the {\it separator}.

The nodal lines of Rabi oscillations are defined based on
Eq.~\ref{EQZ} such that
\begin{eqnarray}
 1-R = x\cdot(1-2n)\sqrt{\frac{\pi}{2}} \, ,\ n\in Z ,
 \mylabel{EQN}
\end{eqnarray}
where $n$ indicates a particular node and
$R$, defined via Eqs.~\ref{EQ20a}-\ref{EQ20}, is 
a function of coordinates $x$ and $y$ defined in Eq.~\ref{EQxy}
such that,
\begin{eqnarray}
& R=2y\cdot\frac{\sqrt{\xi^{2}+1}+1}{2\xi} \,
e^{-\frac{\sqrt{\xi^{2}+1}-1}{2\xi}} \, ,
\mylabel{EQR}
\end{eqnarray}
where
\begin{eqnarray}
& \xi = \frac{4\delta_\omega}{\Gamma}\ \frac{y}{r}\ .
\mylabel{EQxi}
\end{eqnarray}
$R$ can be also expressed using a power series of $\xi$ such that
\begin{eqnarray}
R=\frac{y}{\sqrt{e}}\,
\(1 + \frac{3}{2}\xi^{-1} + \frac{7}{8}\xi^{-2}+ {\cal O}(\xi^{-3}) \) ,
\mylabel{EQE}
\end{eqnarray}
which we will use below to explore the effect of
 magnitude of the pulse frequency width $\delta_\omega$ on the studied phenomenon.

\begin{figure*}
	\begin{center}
		\includegraphics[width=5in]{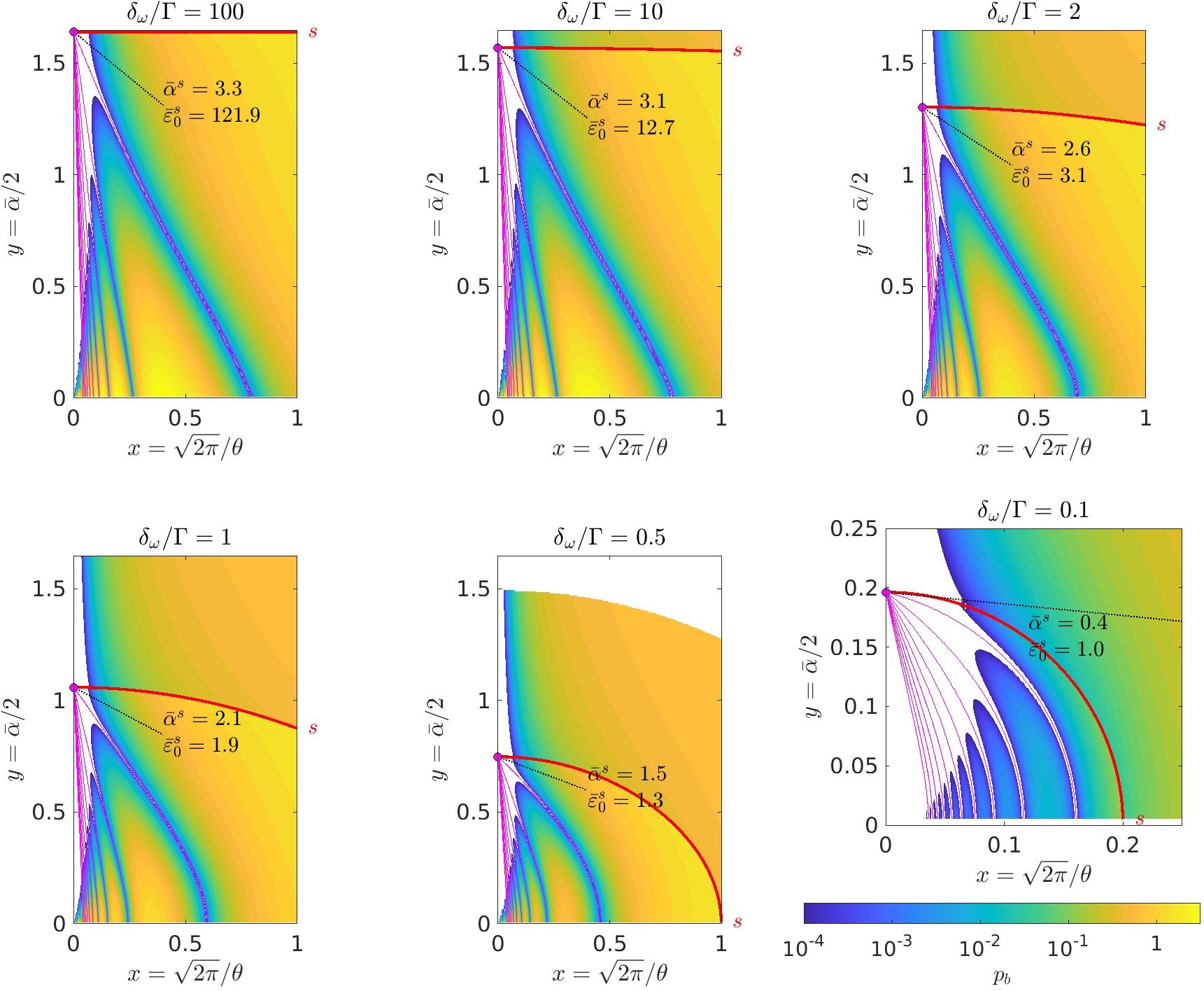}
	\end{center}
	\caption{
		This figure demonstrates the intriguing phenomenon that
		zeros of Rabi oscillations (marked by magenta curves online)
		converge and cross in a single point 
		if the frequency width $\delta_\omega$
		of the chirped pulses is fixed.
		The crossing point also marks a critical change of behavior,
		as it lies on the separator $s$ between
		the regions of Rabi oscillations (below $s$) and
		rapid adiabatic passage (above $s$).
		{\it Note that as the ratio between 
			the pulse frequency width and the resonance width
			is varied, the crossing point is represented by
			different points $[\bar\alpha^s,\bar \varepsilon_0^s]$
			on the separator $s$ (compare Fig.~2)}.
		This characteristic behavior of the zeros
		is proposed as a suitable experimental
		means to determine the values $[\bar\alpha^s,\varepsilon_0^s]$
		by an extrapolation of the curves of the zeros
		starting at accessible regions $p_b>0.001$ to
		$x=0$.		
		Here, the survival probability $p_b$ is displayed as a function
		of the reciprocal pulse area ($x$-axis, $x=\sqrt{2\pi}\hbar/\theta$) 
		and effective
		chirp ($y$-axis, $y=\bar\alpha/2$), where
		the values of the survival probability are indicated by the
		shadow-scale (color-scale online) displayed in the 
		bottom right.
		The value of
		the frequency width $\delta_\omega$
		of the pulse is provided above the figures where we give
		its relative value with respect
		to the resonance width $\Gamma$ of the excited state.	
	}
	\mylabel{FigFrame2}
\end{figure*}

The first panel in Fig.~\ref{FigFrame2} illustrates the situation
where the pulse frequency width $\delta_\omega$
is far larger then the resonance width $\Gamma$, namely the ratio of $\delta_\omega/\Gamma$ is as high as $100$.
This special case represents the limit $\xi\to\infty$, see Eqs.~\ref{EQxi}
and \ref{EQE},
where
\begin{eqnarray}
R\approx \frac{y}{\sqrt{e}} .
\end{eqnarray}
By substitution of this limiting expression for $R$ into Eq.~\ref{EQN}
we obtain equations for different nodes $n$ in the $[x,y]$ plane such that
\begin{eqnarray}
y = \sqrt{e} - x\cdot (2n-1)\ \sqrt{\frac{\pi e}{2}}  \ , \ n\ge 1 \ .
\end{eqnarray}
The fact that $y$ is linearly dependent on $x$ is in agreement with the shape of nodes shown in the first panel in Fig.~\ref{FigFrame2}
The curves for different nodes, $y^{(n)}$, intersect at the point $[x=0,y^{(n)}=y_0=\sqrt{e}]$. 
%This value corresponds with the
%chirp $\bar \alpha = 2\sqrt{e}$, which is the critical
%chirp in the Hermitian limit.

\def\f{\frac{\Gamma}{\delta_\omega}}
Next panel in Fig.~\ref{FigFrame2} displays the situation where
the ratio $\delta_\omega/\Gamma$ is still relatively high
being equal to $10$.
The nodes are still linear but the point of intersection 
is clearly shifted to a smaller value, $y_0<\sqrt{e}$.
The shift of $y_0$ from the asymptotic value
$\sqrt{e}$ can be derived using the first
order expansion of $R$ based on Eqs.~\ref{EQE}
and \ref{EQxi} which is given by,
\begin{eqnarray}
	R\approx\frac{y}{\sqrt{e}}\,
	\(1 + \frac{3}{2}\xi^{-1}\) 
	= \frac{y}{\sqrt{e}}\,
	+ \frac{r}{\sqrt{e}}\,
	\frac{3\Gamma}{8\delta_\omega} \approx \frac{y}{\sqrt{e}}
	\(1 + \frac{3\Gamma}{8\delta_\omega}\) .
\end{eqnarray}
This expression for $R$ is substituted to Eq.~\ref{EQN} which defines
the nodes and we obtain
\begin{eqnarray}
	y \approx
	\frac{\sqrt{e} - x\cdot(1-2n)\sqrt{\frac{\pi e}{2}}}{1 + \frac{3\Gamma}{8\delta_\omega}} 
\end{eqnarray}
which shows that the nodal lines in the $[x,y]$ plane
remain linear within the first order approximation of $R$.   
From here we obtain also the negative shift of the
crossing point $y_0$ by setting $x=0$ such that
\begin{eqnarray}
	y_0 \approx
	{\sqrt{e}}\({1 + \frac{3\Gamma}{8\delta_\omega}}\)^{-1} < \sqrt{e} .
	\mylabel{eq33}
\end{eqnarray}

As the ratio $\Gamma/\delta_\omega$ is even larger, see the 
third to sixth panels in Fig.~\ref{FigFrame2}, 
the nodal lines deviate from  linearity but we can derive
the meaning of the intersecting point $y_0$ in this general
case simply by substitution of $x=0$ to Eq.~\ref{EQN}. 
Eq.~\ref{EQN} shows immediately the fact that nodal curves intersect at criticality, i.e. for $R=1$, compare Eq.~\ref{EQC} and the text along with.
The value of $y_0$ in the intersection $[x=0,y=y_0]$ is obtained 
from the condition $R=1$ applied to Eq.~\ref{EQR}, where we substitute
$y=r$ in the definition of $\xi$ in Eq.~\ref{EQxi}.
We find,
\begin{eqnarray}
	& y_0=\frac{\xi}{\sqrt{\xi^{2}+1}+1} \,
	e^{\frac{\sqrt{\xi^{2}+1}-1}{2\xi}} \, ,\quad
	& \xi = \frac{4\delta_\omega}{\Gamma}\ .
\end{eqnarray}

Let us summarize; we have just proved that nodal planes of the oscillatory
structure in the three-dimensional
space of pulse parameters given by temporary pulse area,
effective chirp, and frequency width of the pulse, or any other
equivalent set of three parameters, collapse into the separator $s$
in the infinite temporal pulse area limit $\theta\to\infty$.
This remarkable phenomenon is apparent for survival probability $p_b$
being shown as a function of reciprocal temporal pulse area ($x=\sqrt{2\pi}/{\theta}$) and effective chirp ($y=\bar \alpha/2$)
when frequency width of the pulse $\delta_\omega$ is held fixed as shown in Fig.~\ref{FigFrame2}.
Note that nodal lines of the oscillating survival probability
can be found experimentally in this frame, as long as 
amplitudes of $p_b$ are relatively large in the major part of the $[x,y]$
space, see the scale given below the last frame in Fig.~\ref{FigFrame2}.
We refer to this fact when giving our suggestions for experiment below.

\subsection{Experimental proposal to localize the critical change of behavior -- separator (Step 1)\mylabel{S33}}

\noindent
Let us explain {\it how this phenomenon can be applied experimentally}.
The goal would be to measure the point $[x=0,y=y_0]$ experimentally
and by knowing that this point is a part of the separator $s$, 
find the corresponding pulse parameters in the
$[\bar\alpha,\bar\varepsilon_0]$ plane which
are related to $y_0$ through Eqs.~\ref{EQxy} and \ref{EQr} such that
\begin{eqnarray}
	&\bar{\alpha}^s = 2 y_0, \quad
	&\bar \varepsilon_0^s = \frac{2\delta_\omega}{\Gamma} \cdot \frac{1}{y_0}\ .
\end{eqnarray}

Up to now we viewed the problem within the reduced frame, see Eqs.~\ref{barpar}, which is also used to define the $[x,y]$ plane.
In a practical experimental setup, however, 
the reduced pulse parameters cannot
be applied as long as the transition moment $\mu$
is considered unknown. Therefore system dependent 
parameters $[x_\mu, y_\mu]$ will be used instead of $[x,y]$
such that
\begin{eqnarray}
	& x_\mu = x\, \mu \equiv \frac{\hbar}{\tau\varepsilon_0^{max}},\quad
	& y_\mu = y\, \mu \equiv \frac{2\hbar\alpha\tau}{\varepsilon_0^{max}}\ .
\end{eqnarray}
Note that $x_\mu$ and $y_\mu$
depend on the usual pulse parameters such as peak amplitude
$\varepsilon_0^{max}$, pulse length $\tau$, and pulse chirp $\alpha$,
Eq.~\ref{EQ2}.

It is assumed that survival probability $p_b$ will be measured
throughout the parametric plane $[x_\mu, y_\mu]$ while
frequency width of the pulse $\delta_\omega$ will be held fixed.
In a realistic experimental setup, chirped laser pulses are
often designed by using grating mirrors applied to an original
non-chirped pulse, where as a result the frequency spread of the pulse
remains the same but the pulse
length and peak amplitude are modified. 
Thus $\delta_\omega$ will be naturally defined
by the pulse length prior to its chirping, while $x_\mu$ and
$y_\mu$ will be varied via the chirp magnitude and field amplitude. 

Nodal lines measured in $[x_\mu, y_\mu]$ for a given fixed value
$\delta_\omega$ will allow to 
determine the crossing point at $[x_\mu=0, y_{0\mu}]$.
Using different values of the pulse frequency width $\delta_\omega$,
the function  $y_{0\mu}(\delta_\omega)$ shall be obtained from the
spectroscopic measurements.
From here the separator $s$ can be obtained up to the system
dependent scaling via the transition moment $\mu$.
The separator is measured in the following form,
\begin{eqnarray}
	&\bar{\alpha}^{(s)}_\mu(\delta_\omega) = 2 \, y_{0\mu}(\delta_\omega),
	\quad
	&\varepsilon_0^{max(s)} = \frac{\delta_\omega}{y_{0\mu}(\delta_\omega)}\ ,
	\mylabel{eq37}
\end{eqnarray}
where $\bar\alpha_\mu$ is a system dependent pulse parameter
related to the effective chirp $\bar\alpha$
such that
\begin{equation}
	\bar{\alpha}_\mu = \bar{\alpha} \mu = 2\hbar \,  \frac{\alpha\tau}{\varepsilon_0^{max}}\ .
\end{equation}

Now let us discuss what we expect from this type of measurement.
First, it is expected that the separator would display a curvature
as the laser amplitude gets near exceptional point EP, $\varepsilon_0^{max(s)} \approx
\varepsilon_0^{EP}$, where the separator
should be found bended towards $\bar{\alpha}_s\to 0$  exactly as predicted in Fig.~2. 
Bending of the separator is associated with a small ratio between 
the pulse frequency width $\delta_\omega$ and resonance width $\Gamma$,
$\delta_\omega/\Gamma\le1$,
which is shown in the lower panels in Fig.~\ref{FigFrame2}.
On the opposite side, as $\delta_\omega/\Gamma>1$, 
$y_{0\mu}$ approaches the asymptotic value of 
\begin{equation}
	y_{0\mu} \to \mu \sqrt{e} 
\end{equation}
from below, see Eq.~\ref{eq33} and upper panels of Fig.~\ref{FigFrame2}.
This regime corresponds with the asymptotic part of the
separator $s$, where variations of $\varepsilon_0^{max(s)}$  (as long as $\varepsilon_0^{max(s)} \gg \varepsilon_0^{EP}$)
do not affect the critical chirp,
$\bar\alpha_\mu^{(s)}\approx const.$, see Fig.~2.

An experimental verification of this indicated {\it general shape of the separator}
would provide clear fingerprints  of
time-symmetric dynamical encircling of EP in frequency--amplitude domain as
shown in Fig.~1.

\subsection{Method for experimental verification of the asymptotes of the separator (Step 2)}

\noindent
Up to now we have shown how the separator can be measured, while
the weight of our argument lies on agreement 
between a measured and the predicted {\it shape} of the separator.
An additional experimental approval of the theoretical predictions
regarding the separator $s$ would represent a
verification of its asymptotes 
for $\bar \alpha_\mu \to 0$ and $\varepsilon_0^{max(s)}\to \infty$.
In the theoretical prediction they correspond with the 
quantities  $\varepsilon_0^{EP}$ and $\mu\sqrt{e}$, respectively,
which calls for a verification using another experimental method.

Let us propose a measurement of Rabi oscillations where the
chirp is not present as such an independent method.
The oscillations of survival probability $p_b$ in Eq.~\ref{EQ9} are
given by the term $Z$ for which we derived the approximate form
Eq.~\ref{EQZ}. Now, as the chirp is not present, $R$ is replaced
by $X=1/\bar\varepsilon_0$, see Eqs.~\ref{EQ20a}-\ref{EQ20}. 
The oscillatory term $Z$ then reads such that
\begin{equation}
	Z \propto \cos^2 \[\frac{\theta}{2\hbar} \(1/\bar\varepsilon_0 - 1\)\] .
\end{equation}
From here, the zeros of $Z$ (corresponding to zeros of
survival probability $p_b$ measured for zero chirp) are defined as
\begin{equation}
\theta = (2n+1)\ 
	\frac{\pi\hbar}{1 - 1/\bar\varepsilon_0 }
\end{equation}
Let us substitute for the relative variables
$\theta$ and $\bar\varepsilon_0$ using realistic
pulse parameters, Eqs.~\ref{barpar}. We obtain
a theoretical prediction for the zeros on the survival probability
$p_b$
given by,
\begin{equation}
	\varepsilon_0^{max}(n) = \varepsilon_0^{EP} + (2n+1)\ 
	\frac{\sqrt{2\pi}\hbar}{2\tau\mu},\quad
	\varepsilon_0^{max}(n) \ge \varepsilon_0^{EP} ,
	\mylabel{EQshift}
\end{equation}
assuming that the pulse length $\tau$ is fixed while
field amplitude $\varepsilon_0$ is varied. Note again that
Eq.~\ref{EQshift} represents a theoretical prediction for
Rabi oscillations when EP is {\it encircled}, i.e. 
the field amplitude $\varepsilon_0^{max}$ must be beyond 
$\varepsilon_0^{EP}$, see Fig.~1.

We propose to use
Eq.~\ref{EQshift} first to determine transition
dipole moment $\mu$ from experimental measurement of the periods
of Rabi oscillations such that
\begin{equation}
	\mu = \frac{\sqrt{2\pi}\, \hbar}{\tau\Delta_{\varepsilon_0^{max}}},
	\quad
	\Delta_{\varepsilon_0^{max}}= \varepsilon_0^{max}(n+1) - \varepsilon_0^{max}(n)\, .
\end{equation}
Using Eqs.~\ref{barpar} one can prove
that the period $\Delta_{\varepsilon_0^{max}}$ corresponds to
the change of temporal pulse area by $\Delta_\theta = 2\pi\hbar$.
This phenomenon has been found for
bound-to-bound transitions (which correspond
with our theory as $\varepsilon_0^{EP}\to0$)
where it has been studied in coherent control~\cite{Vitanov:2001}.
This phenomenon is now found general to bound-to-resonance transitions as well.

Though we do not know the exact equation for the survival probability $p_b$
below $\varepsilon_0^{EP}$, we do know that $p_b$ is {\it not} oscillatory
there.
Therefore the position of the first zero
on $p_b$ is given by
\begin{equation}
	\varepsilon_0^{max}(n=0) = \varepsilon_0^{EP} + 
	\frac{\sqrt{2\pi}\hbar}{2\tau\mu}\ 
\end{equation}
according to Eq.~\ref{EQshift}.
Let us see what this equation says about
the first zero in bound-to-bound transitions where
$\varepsilon_0^{EP}=0$. 
It shows the remaining term $\sqrt{\pi/2}\,\hbar/\tau\mu$ 
which corresponds to the temporal pulse area $\theta=\pi$,
see Eqs.~\ref{barpar}.
Such pulse is used to be called
the $\pi$-pulse in coherent
control, where it is used as a method
 to obtain a maximum transition to the excited state, 
see Ref.~\cite{Vitanov:2001}.
 As we can see immediately,
the first zero of $p_b$ in bound-to-resonance
transitions is shifted exactly by $\varepsilon_0^{EP}$,
i.e. beyond the encircled EP.
 After determining the period of Rabi oscillations in the step
 above one can determine also 
$\varepsilon_0^{EP}$ using this new theoretical prediction.

Here we outlined an independent experimental
method exploring Rabi oscillations
of bound-to-resonance transitions without a chirp 
to determine the transition dipole
moment $\mu$ and the position of the EP $\varepsilon_0^{EP}$.
This method is independent on the measurement of the
separator $s$ described above as there is no search
for the critical change of behavior (Rabi-to-RAP)
in this method.

The two experimental
methods are meant
to be used in conjunction:
First the critical change of behavior 
(Rabi-to-RAP) should be measured using the direct method
described in Section~\ref{S33} where
the general shape and limits of the separator $s$
for  $\bar \alpha_\mu \to 0$ and $\varepsilon_0^{max(s)}\to \infty$ should be determined.
Then these limits, theoretically
predicted as given by the 
quantities  $\varepsilon_0^{EP}$ and $\mu\sqrt{e}$, respectively,
should be compared with measurements 
of $\varepsilon_0^{EP}$ and $\mu$
based on the
Rabi oscillations where chirp is not applied
as described in this Section.

\section{Requirements on suitable atomic systems and laser pulses}

\subsection{Real valued transition moment -- using absorption line profiles}

\noindent
Our above pursued elaborations assumed that the atomic transition moment $\mu$ is real valued, which is assured
for any bound-to-bound transition.
In bound-to-{\it resonance} transitions, the transition dipole moment
is generally complex. It is uniquely defined as~\cite{Pick:2019}
\begin{equation}
	\mu = \sqrt{(1_L|x|2_R)\, (2_L|x|1_R)} \ ,
\end{equation}
where (L) and (R) denotes left and right eigen-vectors~\cite{Moiseyev} 
of the field-free states denoted by numbers (1,2) which are involved in the 
transition.
This definition of transition dipole moment assures symmetry of
the two-by-two Floquet Hamiltonian (Eq.~\ref{EQ1}).

Pick et al.~\cite{Pick:2019} demonstrated that the complex phase of $\mu$
is directly related to a
spectral shift of the resonance absorption line, the so called Fano profile, 
also understood as a result of an interference between 
quasi-bound state and free-particle scattering amplitudes~\cite{Fano:1961}.
We reassume here the corresponding
asymmetric shifted absorption line profile
\begin{equation}
	S(\omega) \propto
	\frac{(\omega-\omega_r + q\, \Gamma/2)^2}
	{(\omega-\omega_r)^2 + (\Gamma/2)^2}\ ,
\end{equation}
where $q$ is an asymmetry parameter.
For $q=0$ a symmetric Lorenzian profile is obtained with maximum
for $\omega=\omega_r$. Otherwise the absorption peak is shifted from $\omega_r$,
while the line profile is asymmetric.
Based on results of Ref.~\cite{Pick:2019},
$q$ is directly related to the complex phase of the
transition dipole moment $\mu$ such that
\begin{equation}
	q = \frac{\cos \phi - 1}{\sin \phi} \approx \phi,
	\quad
	\phi = \arctan \frac{\im \mu^2}{\re \mu^2}\ .
\end{equation}

This excursion 
indicates an interrelation between the symmetry of absorption line profiles
and obtaining {\it time-symmetric EP encircling} using 
Gaussian chirped pulses. It leaves us with a simple
experimental method to determine whether
a given bound-to-resonance transition would demonstrate the Rabi-to-RAP 
switch or not.
Such an experimental tool would be particularly useful in situations where
precise non-Hermitian theoretical calculations of transition dipole moments
do not exist (which is still the case even for larger atoms).

\subsection{Isolated two level system -- a full dimensional numerical simulation for the helium atom}
\def\He{{\rm He}}
\def\s{{\rm s}}
\def\p{{\rm p}}

\noindent
Physical relevance of the studied phenomenon of the switch between Rabi
oscillations and RAP, 
relies on the assumption that the two coupled levels are sufficiently
decoupled from the rest of atomic levels. 
Allow us therefore approve the proposed phenomenon in a realistic system
for which direct (numerically exact) non-perturbative computations of $p_b$
may be simply performed.

We choose a specific class of bound-to-resonance transitions
in the helium atom, which is derived from the ionic bound-to-bound transitions,
\begin{eqnarray}
	\He^+(1\s) (S) \to (\He^+)^*(2\p) (P),
\end{eqnarray}
where instead of the ion we assume a neutral Rydberg atom, where the electron is added to
a Rydberg orbital ($n\p$):
\begin{eqnarray}
	\He^*(1\s n\p) (P) \to (\He)^{**}(2\p n\p) (S,D).
\end{eqnarray}
The doubly excited atom is an autoionization resonance which would
spontaneously decay to the ionized atom:
\begin{eqnarray}
	(\He)^{**}(2\p n\p) (S,D) \to \He^+(1\s) (S) + e^- .
\end{eqnarray}
These transitions are characterized by almost real valued
transition dipole moment. We pick up  
the
transition $\He^* (1\s2\p)$ $\to$ $\He^{**} (2\p^2)$ where the transition dipole moment $\mu$ is given by $\mu=0.763643+0.0001i$.

Numerical method
for quantum dynamics of the helium atom in the presence of
chirped laser pulse has been described in Refs.~\cite{Kapralova-Zdanska:2013,Kapralova-Zdanska:2014},
here we present particular parameters for the simulations,
namely 
184 field-free states of helium forming the basis set included the bound states, quasi-discrete
continuum (He$^+(1s)$ + $e^-$), and the resonances above the first ionization threshold (see Ref.~\cite{ISI:000313642500011}).
The diameter of the space sampled using the ExTG5G primitive Gaussian basis set reached
up to 40\AA. The complex scaling parameter was set to $\vartheta=0.4$.

The initial state for the simulation is represented by the $\He^* \ ^oP\,(1s2p)$ state.
This state is coupled by the XUV pulse with the central wavelength $\lambda = 30.322$~nm
with the doubly excited resonance $\He^{**} \ ^eS\,(2p^2)$ with the lifetime of $\tau=0.11$~ps.
The results of the multidimensional simulations displayed in Fig.~\ref{FigureFullDim} 
represent an evidence that Eqs.~\ref{EQ9}-\ref{EQ10} are valid to a good approximation. 
The sought effect, where
the oscillatory behavior changes due to the coalescence of the TPs in the complex plane
of adiabatic time, is clearly present in the full-dimensional case.

\begin{figure}
	\begin{center}
		\includegraphics[width=2.25 in]{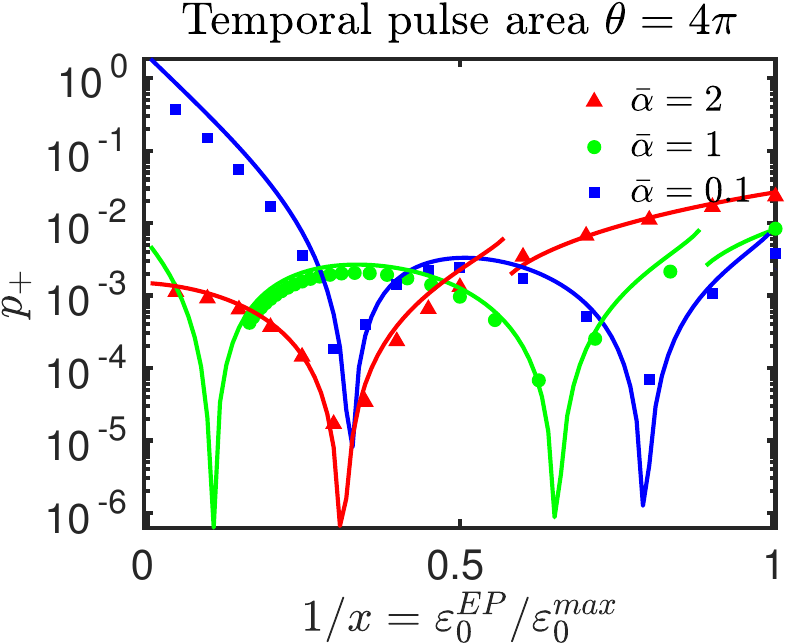}
		\includegraphics[width=2.25 in]{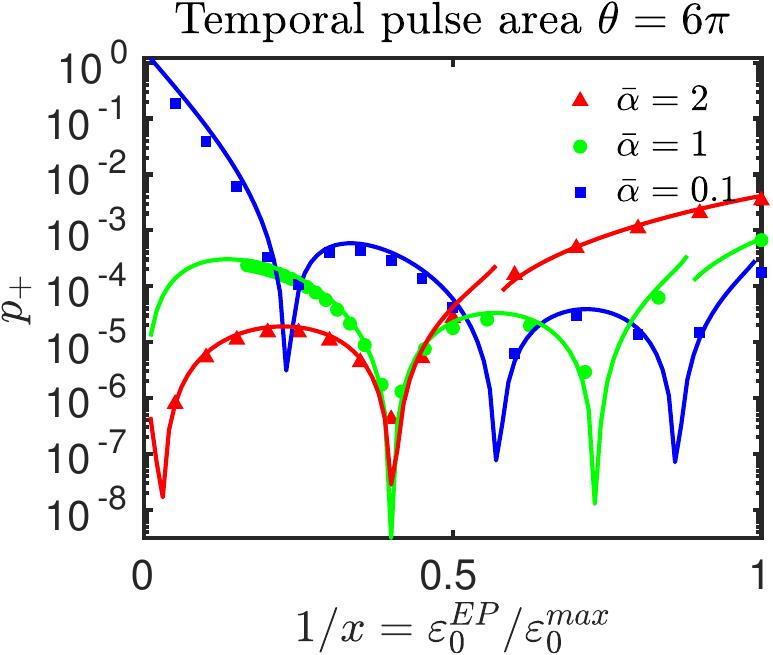}
		\includegraphics[width=2.25 in]{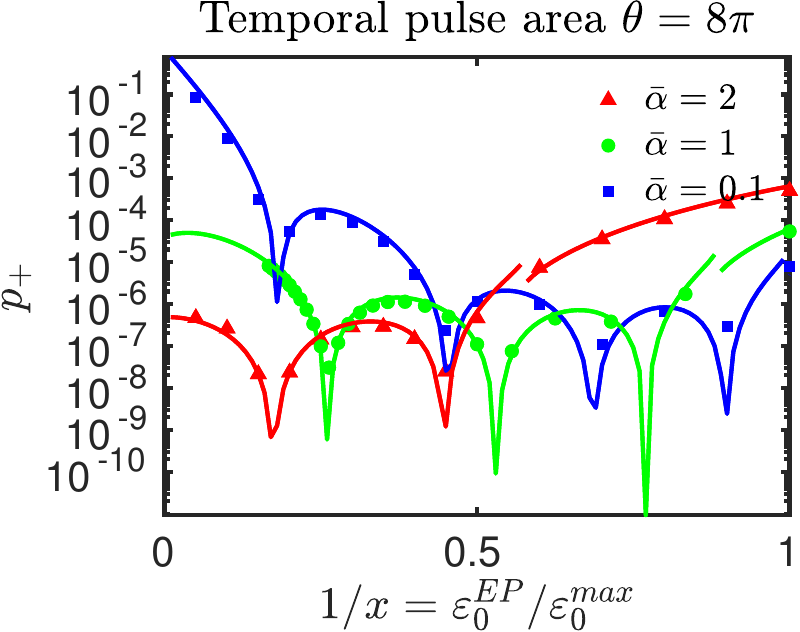}
	\end{center}
	\caption{
		These figures demonstrate the qualitative and even semi-quantitative
		aplicability of the presented analytical solution obtained
		using a number of approximations, starting from the two level approximation through the first order
		perturbation method down to the long pulse limit, to a real atomic system, which is represented here
		by a full dimensional simulation of the helium atom interacting with the laser field.
		The symbols ($\blacktriangle$, $\bullet$, $\blacksquare$) represent results from numerical simulations
		using a basis set of 184 field-free helium states, which are coupled by the
		interaction with the classically approximated field.
	}
	\mylabel{FigureFullDim}
\end{figure}

Let us briefly comment also on the experimental feasibility of the predicted phenomenon
in the helium atom.
The pulse peak intensity reaches up to 4~\TWcm\  and the pulse length reaches up to 5~ps in this
case, where these quantities are correlated to obtain the particular laser parameters 
$\bar \alpha,\bar\varepsilon_0,\theta$.
Such pulses are generally available today.
Below we add a discussion of precision demands for frequency, which are
high in this case of XUV range.

\subsection{Precision demands on the carrier frequency of the Gaussian chirps}

\noindent
The presented phenomenon requires the mean laser frequency
being equal to the real energy difference between the two involved 
field free states.
Here we provide a numerical simulation of the system
including a static detuning of the laser frequency.
The pulse would be modified such that
\begin{eqnarray}
& \varepsilon_0(s\tau) = \frac{2\mu}{\Gamma}\,\x\, e^{-s^2/2}, \cr
&\omega(s\tau) = \omega_r + \Delta_\omega
+ \balpha\, \x \cdot \frac{s}{2} \cdot \frac{\mu^2}{\hbar\Gamma} \ ,
\end{eqnarray}
where $\Delta_\omega$ is a static detuning.
If $\Delta_\omega$ is non-zero then the time-symmetry 
of the energy split is broken.
This fact implies that the opposite TPs in the complex
adiabatic-time plane (Fig.~3) are shifted from their
original symmetrical layout and therefore the interference
phenomenon which is responsible for Rabi oscillations 
is less efficient, i.e. the oscillatory minima are not
exact zeros. An impact of asymmetry due to
static detuning is illustrated in Fig.~\ref{FigDetun}
which has been obtained by solving Eq.~\ref{EQ6} numerically for 
detuned laser pulses. 
\begin{figure}[!h]
	\begin{center}
		\includegraphics[width=2.5in]{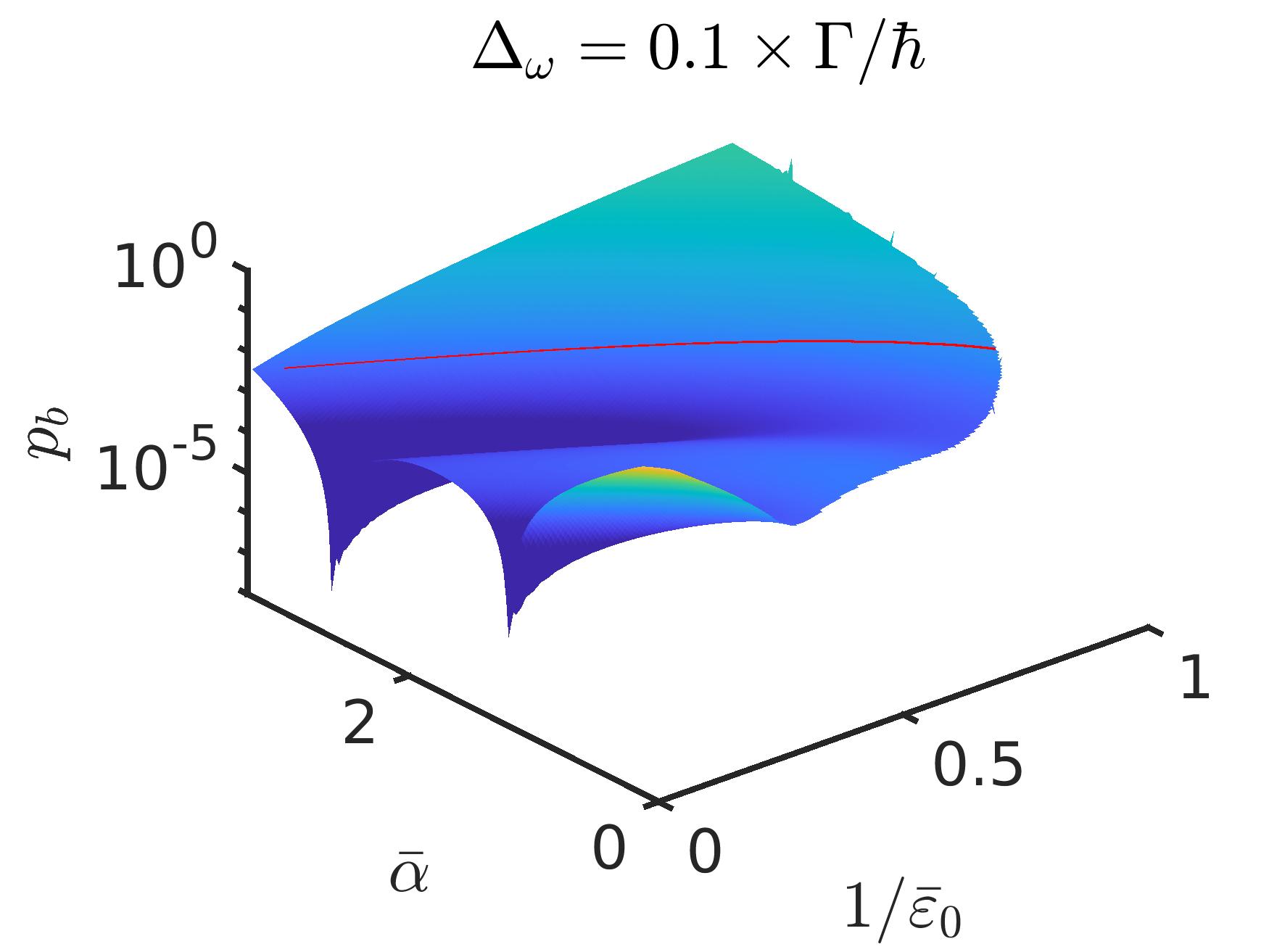}
		\includegraphics[width=2.5in]{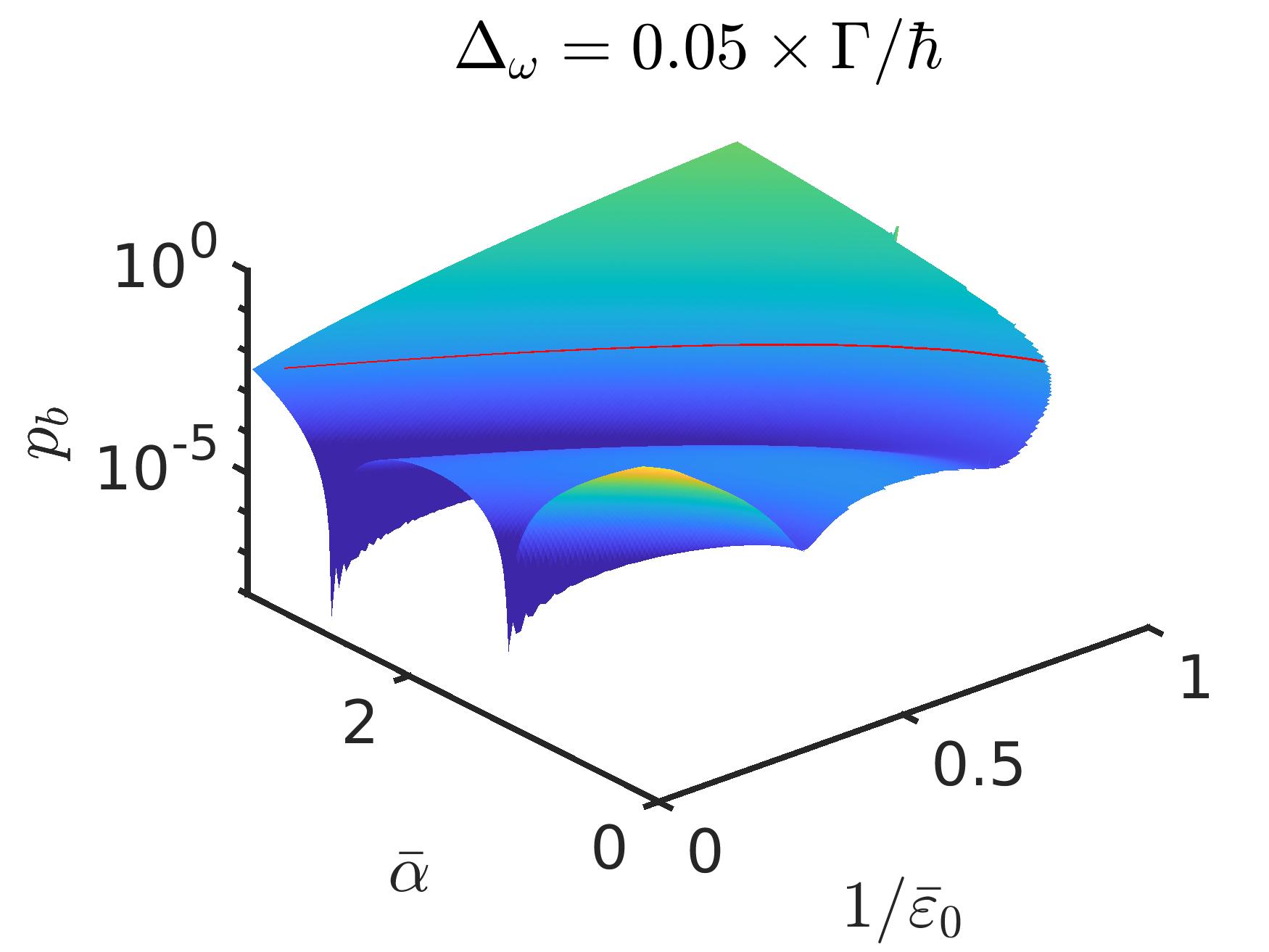}
		\includegraphics[width=2.5in]{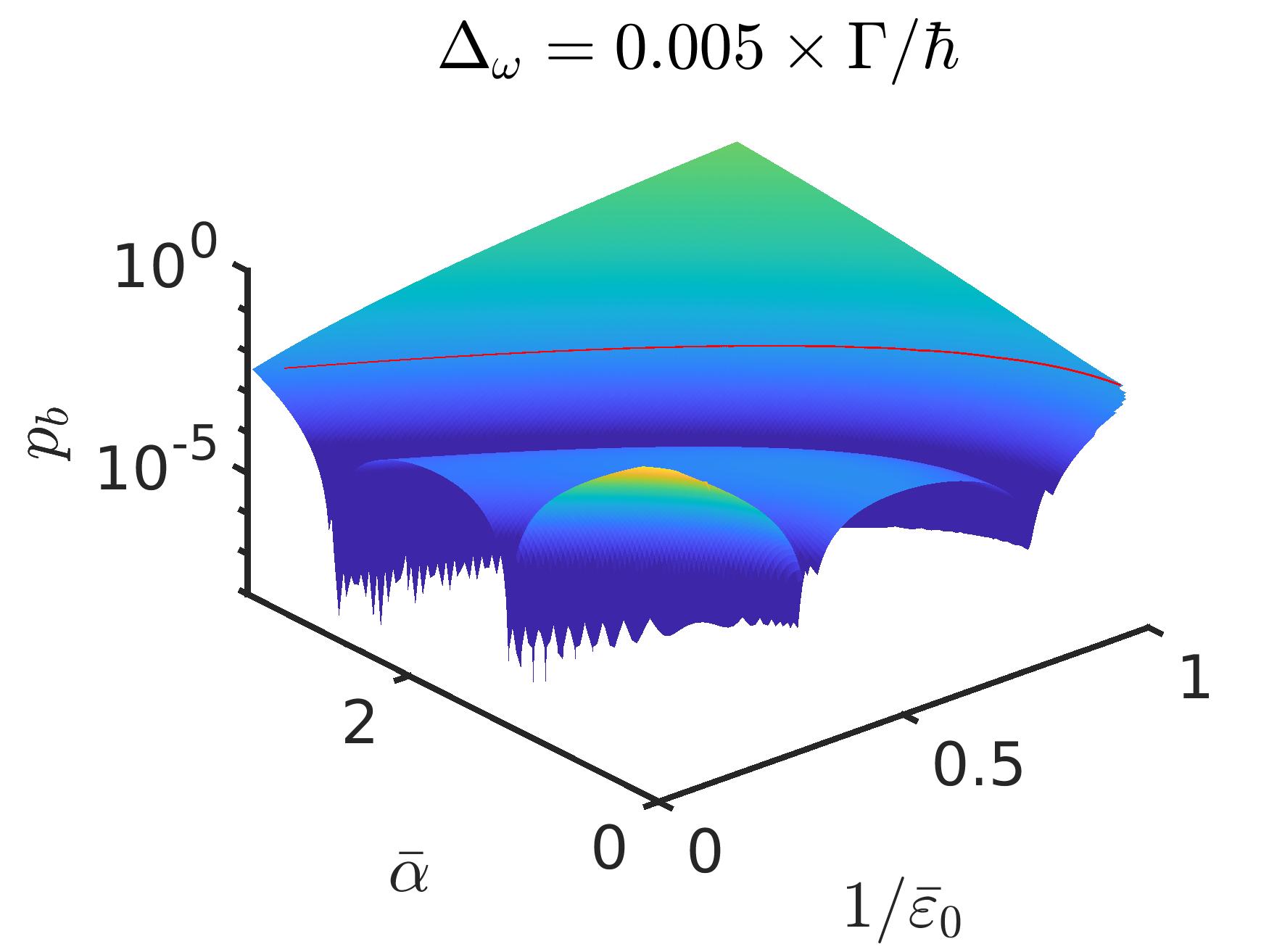}
	\end{center}
	\caption{
		A static detuning removes the time-symmetry of the energy split.
		As a result of this, nodes of the oscillatory regime 
		are less pronounced when the effective chirp $\balpha$
		is small and the field amplitude is close to the encircled EP,
		$1/\bar\varepsilon_0 \to 1$.
		This effect is demonstrated here for the pulse area $\theta=4\pi$. We
		show the results of the first order perturbation approximation.
	}
	\mylabel{FigDetun}
\end{figure}

Fig.~\ref{FigDetun} shows that nodes of the Rabi oscillations
are not so well pronounced but at the same time they stay in place
(are not shifted)
in the $1/\bar\varepsilon_0$--$\balpha$ plane as a detuning is introduced.
From the experimental point of view this means that 
there is a certain tolerance for the precision of the 
carrier frequency $\omega_r$. Based on simulations which are
presented in Fig.~\ref{FigDetun} we define the tolerance factor $f_p$
which defines the detuning with respect to the resonance lifetime
such that
\begin{equation}
	\Delta_\omega = f_p \ \Gamma/\hbar ,
\end{equation}
where $f_p$ represents the factors $0.01$, $0.1$, and $0.2$ in the
panels of Fig.~\ref{FigDetun}. Let us consider the value of $f_p=0.1$ 
as the precision requirement.
When this requirement 
is put in terms of wavelength, the demand on wavelength
precision is given by
\begin{eqnarray}
\Delta_\lambda = f_p \cdot \frac{\lambda^2\,\Gamma}{4\pi\hbar c } ,
\end{eqnarray}

According to this formula, for the helium transition which was suggested above, the
detuning in a experiment should not exceed $\Delta_\lambda = 0.4$~pm.
In other systems where the wavelength of transitions would be in the IR
range such as 800~nm, 
achieving the Rabi-to-RAP phenomenon
would be possible with much lesser precision of the carrier frequency
 of $0.1$~nm when about the same resonance
width $\Gamma$ is assumed.

\def\x{\bar \varepsilon_0}
\def\balpha{\bar\alpha}

\section{Conclusions}

\noindent
We have shown that, in particular atomic bound-to-resonance transitions,
the dynamical encircling of an EP is manifested by a critical switch between oscillatory
 and monotonic behavior of the occupancy $p_b$ of the initial atomic bound state.
We rationalize this phenomenon theoretically as a consequence of
coalescence of TPs (branchpoints of the adiabatic Hamiltonian in complex time
plane). Such an intriguing interconnection between EPs and TPs has not yet been discussed
in this context of laser driven atoms. 

The studied critical phenomenon lends itself well to
an experimental realization in driven atoms using linearly chirped Gaussian pulses.
We propose a suitable experimental
method how the critical change of behavior can be 
directly localized
 in the parameter space of the linearly chirped Gaussian pulses.
 The critical change of behavior defines a typical separating line 
 between the oscillatory and monotonic areas in the pulse parameter space.
The starting point of the separator in the pulse parameter space 
is identical with the laser amplitude associated with the EP in the
encircling (frequency--laser amplitude) plane.

Experimental demonstration of the presented phenomenon
would represent a novel way how to
detect fingerprints of the encircled EP in laser driven atoms and
even allow for a direct measurement of the EP position in the
encircling space.

\ack

\noindent
The work was financially supported in parts by the Czech Ministry of Education,
Youth and Sports (Grant LTT17015),
the Grant Agency of the Czech Republic (Grant 20-21179S)
 (P.R.K. and M.S.), 
and by the Israel Science Foundation grant No.~1661/19 (N.M.).

\appendix

\section{Linearly chirped Gaussian pulses}

\subsection{\label{APchirp1}Gaussian pulse chirping based on group velocity dispersion}

\noindent
The practival way of constructing chirped pulse is by using 
a frequency decomposition of a non-chirped laser pulse, where individual
frequencies go through a different path length, after which they are
assembled back.
The difference between the length of paths for frequencies is
given by the {\it group velocity dispersion}.
The relation between the group velocity dispersion and the chirp
is derived here.

Let us define the laser pulse before chirping such that
\begin{equation}
	f(t) = \varepsilon_0^{max} \exp\( -{t^2\over{2\tau_0^2}} + i\omega_0 t\),
	\mylabel{EQA1}
\end{equation}
where $\omega_0$ represents the carrier frequency, and $\tau_0$ represents
the pulse length before chirping.
By applying the Fourier transform of $f(t)$ we obtain the frequency distribution
of such a pulse given by
\begin{equation}
	\tilde f(\omega) = {\varepsilon_0^{max} \over \delta_\omega} 
	\exp\[ -{(\omega-\omega_0)^2\over{2\delta_\omega^2}} \],
\end{equation}
where $\delta_\omega$ represents the frequency width related to the pulse length
$\tau_0$ such that
\begin{equation}
	\delta_\omega = {\hbar \over \tau_0} .
	\mylabel{EQA3}
\end{equation}
The same pulse can be also defined in the mixed
time-energy domain via the Wigner distribution~\cite{Wigner32,Wigner84,Lee}
(as a Fourier transform over the density matrix $f^*(t-t'/2)f(t+t'/2)$
over the correlation coordinate $t'$)
such that
\begin{equation}
	f_w(\omega,t) = {(\varepsilon_0^{max})^2 \tau_0\over\sqrt{\pi}}
	\exp\[-\frac{t^2}{\tau_0^2} - \frac{(\omega-\omega_0)^2}{\delta_\omega^2}\] .
	\mylabel{EQA4}
\end{equation}
General properties of the Wigner distribution~\cite{Lee} assure that
\begin{eqnarray}
	& |f(t)|^2 = \int d\omega\, f_w(\omega,t), \label{EQA5} \\
	& |\tilde f(\omega)|^2 = \int dt\, f_w(\omega,t)\, .
	\mylabel{EQA6}
\end{eqnarray}

The practical way how the chirping is usually done is by imposing the time shift
between different frequency components. 
Mathematically this means that the time $t$ is replaced by a function
related to the frequency component $\omega$ via the group dispersion velocity $\nu_{GDV}$,
\begin{equation}
	t \to t + \nu_{GDV} (\omega-\omega_0),
\end{equation}
in Eq.~\ref{EQA4}. The Wigner distribution for the chirped pulse $g_w$ is
then obtained,
\begin{eqnarray}
	&g_w(\omega,t) = {(\varepsilon_0^{max})^2 \tau_0\over\sqrt{\pi}} \cr
	&\times
	\exp\[-\frac{\[t + \nu_{GDV}(\omega-\omega_0)\]^2}{\tau_0^2} - \frac{(\omega-\omega_0)^2}{\delta_\omega^2}\] .
	\cr
	\mylabel{EQA8}
\end{eqnarray}
The chirping does not change the frequency profile of the pulse
as one could prove by applying Eq.~\ref{EQA6} to the definition of $g_w$ in Eq.~\ref{EQA8}.
The time-profile, on the other hand, is modified which one can see by 
applying a few algebraical steps on Eq.~\ref{EQA8} such that:
\begin{eqnarray}
	&g_w(\omega,t) = {(\varepsilon_0^{max})^2 \tau_0\over\sqrt{\pi}} \cr
	&\times
	\exp\[-\frac{t^2}{\tau_0^2 \gamma^2} + \gamma^2\frac{\[\omega 
		- \(\omega_0+\frac{\nu_{GVD}}{\gamma^2\tau_0^4}t\)\]^2}{\delta_\omega^2}\] \, ,
	\cr
	\mylabel{EQA9}
\end{eqnarray}
where
\begin{equation}
	\gamma^2 = 1 + \frac{\nu_{GVD}^2}{\tau_0^4} \, .
	\mylabel{EQA10}
\end{equation}
The
 time-profile of the chirp remains Gaussian with the new pulse length $\tau$,
\begin{equation}
	\tau = \tau_0 \, \gamma,
	\mylabel{EQA11}
\end{equation}
which one can prove by integrating Eq.~\ref{EQA9} with respect to $\omega$
based on Eq.~\ref{EQA5}.

Next we transform the Wigner distribution $g_w$ back to the complex
definition of the pulse in the time-domain, which will be denoted as $g(t)$
in order to obtain the chirp $\alpha$,
which is defined as the gradient of the complex phase: 
\begin{equation}
	\omega_0 + \alpha \cdot t = \frac{d \arg g(t)}{dt}
	\mylabel{EQA12}
\end{equation}
The complex phase of the original pulse $f(t)$ (given by $\omega_0$, Eq.~\ref{EQA1})
shall differ from that of the new pulse $g(t)$ due to the
modification of the pulse ($f_w \to g_w$) via the group velocity dispersion $\nu_{GVD}$.
The right hand side of Eq.~\ref{EQA12} is readily obtained from
the Wigner distribution as
\begin{equation}
	\frac{d\arg g(t)}{dt} = \int d\omega \, \omega \, g(\omega,t) .
\end{equation}
From here we obtain the linear time dependence defining the chirp,
\begin{equation}
	\alpha = \frac{\nu_{GVD}}{\nu_{GVD}^2 + \tau_0^4} .
	\mylabel{EQA14}
\end{equation}

\subsection{Relations between parameters of Gaussian linearly chirped pulses}

\noindent
The definitions that we derived above, in particular Eqs.~\ref{EQA10}-\ref{EQA11}
and Eq.~\ref{EQA14} help us to derive further relations between different
parameters of the Gaussian linearly chirped pulses.
Eqs.~\ref{EQA10}-\ref{EQA11} are combined such that
\begin{equation}
	\tau^2 \tau_0^2 = \tau_0^4 + \nu_{GVD}^2 .
	\mylabel{EQA15}
\end{equation}
By putting together Eqs.~\ref{EQA15} and \ref{EQA14} we 
obtain a simple relation for
the group velocity dispersion such that
\begin{equation}
	\nu_{GVD} = \tau^2\tau_0^2\, \alpha \, .
	\mylabel{EQA16}
\end{equation}
This definition is substituted back to Eq.~\ref{EQA15} where we obtain
a relation between chirp, pulse length, and the pulse length of the
original pulse such that,
\begin{equation}
	\tau^2 = \tau_0^2\ ( 1 + \tau^4 \alpha^2)\ .
	\mylabel{EQ17}
\end{equation}
The pulse length of the original pulse $\tau_0$ is related to the
frequency width $\delta_\omega$ (Eq.~\ref{EQA3}) which is
identical for the original and chirped pulses, see Eq.~\ref{EQA8}
and the text below.
Thus using Eqs.~\ref{EQ17} and \ref{EQA3} we obtain
a relation between the frequency width  of a chirped pulse
and the pulse length $\tau$ and the chirp $\alpha$ as follows,
\begin{eqnarray}
	\delta_\omega^2 = \hbar^2 \tau^{-2} \( 1+ \alpha^2 \, \tau^4 \) \,  .
	\mylabel{EQA18}
\end{eqnarray}

\subsection{Relation between effective parameters of Gaussian linearly chirped pulse}

\noindent
Effective parameters of Gaussian linearly chirped pulse defined in Eqs.~\ref{barpar}
allowed us to derive general equations 
for phenomena related to the symmetric EP encircling
quantum dynamics of any two level system.
Eq.~\ref{EQA18} can be rewritten in terms of the effective pulse variables 
$\balpha$ and $\theta$ (Eq.~\ref{barpar}) such that
\begin{eqnarray}
	\(
	\frac{\delta_\omega}{\Gamma} \frac{2}{\bar \varepsilon_0}
	\)^2
	=  
	2\pi
	\(\frac{\hbar }{\theta}
	\)^{2}+ \(\frac{\balpha}{2}
	\)^{2} \,  ,
\end{eqnarray}
where we first substituted for $\alpha$ in Eq.~\ref{EQA18},
\begin{equation}
	\alpha = \frac{\balpha}{2}\  \frac{\varepsilon_0^{max}}{\tau} \,
	\frac{\mu}{\hbar}\ ,
\end{equation}
and then for $\tau$ using $\theta$,
\begin{equation}
	\tau = \theta \frac{1}{\varepsilon_0^{max}\mu\sqrt{2\pi}},
\end{equation}
and finally for $\varepsilon_0^{max}$ using 
\begin{equation}
	\varepsilon_0^{max} = \bar\varepsilon_0 \, \frac{\Gamma}{2\mu} \ .
\end{equation}

%\bibliography{jourtab,etg,etges,foumouo,heioniz,helium,helium2,moiseyevnew,zamastil,kapralova,burgers,theory,wigner,hedoubly,laserdyn,laserdyn2,laserdyn3,laserdyn4,laserdyn5,xuvlasers,alon,civis,field,helium3,helium4,excep,excep2,rap,pulsearea}

\end{document}